\documentclass[aps,amsfonts,pra,twocolumn,showpacs]{revtex4-1}
\usepackage{epsfig,amsmath,amssymb,bm,epsf,graphicx,psfrag}
\usepackage[all]{xy}
\usepackage{color}

\def\bra#1{\langle#1\vert}
\def\ket#1{\vert#1\rangle}
\def\ketbra#1{\vert#1\rangle\langle#1\vert}

\def\Longarrow{\protect\@lra}
\def\@lra{\relbar\joinrel\relbar\joinrel\relbar\joinrel%
          \relbar\joinrel\rightarrow}

\newcommand{\bc}{\begin{center}}
\newcommand{\ec}{\end{center}}
\newcommand{\be}{\begin{equation}}
\newcommand{\ee}{\end{equation}}
\newcommand{\bea}{\begin{eqnarray}}
\newcommand{\eea}{\end{eqnarray}}

\newcommand{\ncd}{\newcommand}
\ncd{\QCcns}{$QC_{\cal{C}}$}
\ncd{\QCc}{$QC_{\cal{C}}\;$}

\definecolor{libl}{cmyk}{0.2,0.1,0,0}

\newcommand{\tr}{\mathrm{Tr}}

\begin{document}

\title{Transitions in the quantum computational power}

\author{Tzu-Chieh Wei}
\affiliation{C. N. Yang Institute for Theoretical Physics and the
Department of Physics and Astronomy, State University of New York at
Stony Brook, Stony Brook, NY 11794-3840, USA}
\author{Ying Li}
\affiliation{Centre for Quantum
 Technologies, National University of Singapore, 2 Science Drive 3,
Singapore} \affiliation{Department of Materials, University of
Oxford, Parks Road, Oxford OX1 3PH, United Kingdom}
\author{Leong Chuan Kwek}
\affiliation{Centre for Quantum
 Technologies, National University of Singapore, 2 Science Drive 3,
Singapore} \affiliation{National Institute of Education and
Institute of Advanced Studies, Nanyang Technological University, 1
Nanyang Walk, Singapore}
\date{\today}

\begin{abstract}
We construct two spin models on lattices (both two and
three-dimensional) to study the capability of quantum computational
power as a function of temperature and the system parameter. There
exists a finite region in the phase diagram such that the thermal
equilibrium states are capable of providing a universal
fault-tolerant resource for measurement-based quantum computation.
Moreover, in such a region the thermal resource states on the 3D
lattices can enable topological protection for quantum computation.
The two models behave similarly in terms of quantum computational
power. However, they have different properties in terms of the usual
phase transitions. The first model has a first-order phase
transition only at zero temperature whereas there is no transition
at all in the second model. Interestingly, the transition in the
quantum computational power does not coincide with the phase
transition in the first model.
\end{abstract}

 \maketitle

\section{Introduction}
Transitions in phases of matter, such as melting of ice and boiling
of water, is common in everyday life~\cite{Thermodynamics}. They
also occur in zero temperature, where properties of the system are
governed, instead of thermal effect, by quantum mechanical
fluctuations~\cite{Sachdev}. Tremendous understanding has been
gained on the transitions in phases of matter. Recently, ideas from
quantum information and computation~\cite{NielsenChuang00} give rise
to new perspectives on examining phases of matter, such as
topological phases and their classification~\cite{Wen}. Moreover,
from the viewpoint of computational universality in
measurement-based quantum computation
(MBQC)~\cite{GoChua,NielsenLeungChilds,Oneway,Oneway2,RaussendorfWei12},
a few works have suggested that resource states can emerge from
certain quantum phases of
matter~\cite{DohertyBartlett,Miyake,BartlettBrennenMiyakeRenes,ElseSchwarzBartlettDoherty,ElseBartlettDoherty,FujiiNakataOhzekiMurao}
and that the transition in the quantum computational capability
results in a new notion of phase
transitions~\cite{GrossEisertEtAl,Browne,BarrettBartlettDohertyJenningsRudolph,Darmawan}.

Here, we construct two models to investigate their ground states and
thermal states for
 providing universal quantum computational resource for MBQC.
   As we shall see both models exhibit similar
`phase diagrams' in terms quantum computational power, in both two
and three dimensions. The advantage of 3D offers the possibility of
topological protection in carrying out quantum computation, even at
higher temperatures than in 2D. The two models are natural extension
from a symmetric model that we considered previously~\cite{Thermal},
 and the asymmetric parameter introduced here can be used study its effect on computational universality,
 as well as the possibility
 to tune the system through a quantum phase transition.
 They are exactly
solvable, and thus also allow us to study and compare with the usual
transitions in phases of matter. The first model has a first-order
phase transition only at zero temperature and it does not coincide
with the transition in the quantum computational power. Moreover,
even though there is no phase transition at any finite temperature,
there is a region at finite temperature that supports universal
quantum computation. The second model does not have a phase
transition at zero temperature but has a transition in quantum
computational power at both zero and finite temperatures.

The remaining of the paper is organized as follows. In
Sec.~\ref{sec:models} we introduce the two models, which are defined
on any trivalent lattices either in two or three dimensions. We
focus on the ground-state properties as well as the phase diagram at
finite temperatures. In Sec.~\ref{sec:ground} we discuss
zero-temperature quantum computational capability and show the
existence of a range of the system parameter, where the ground state
can provide a useful resource for universal MBQC. In
Sec.~\ref{sec:Thermal} we turn to the finite temperatures and
consider the thermal effects on quantum computational universality.
We use the techniques of fault-tolerance quantum computation (FTQC)
to map out regions in the phase diagram where FTQC can still be
carried out by using thermal states for the universal MBQC. The
corresponding phase diagrams of quantum computational power are
obtained for both models in both two and three dimensions. It is
worth mentioning that the 3D models provide topological protection
and hence the transition temperature in QC power is higher than that
in 2D. We make concluding remarks in Sec.~\ref{sec:conclude}.
\section{Two model Hamiltonians}
\label{sec:models}
\begin{figure}
\vspace{1cm}
   \includegraphics[width=8cm]{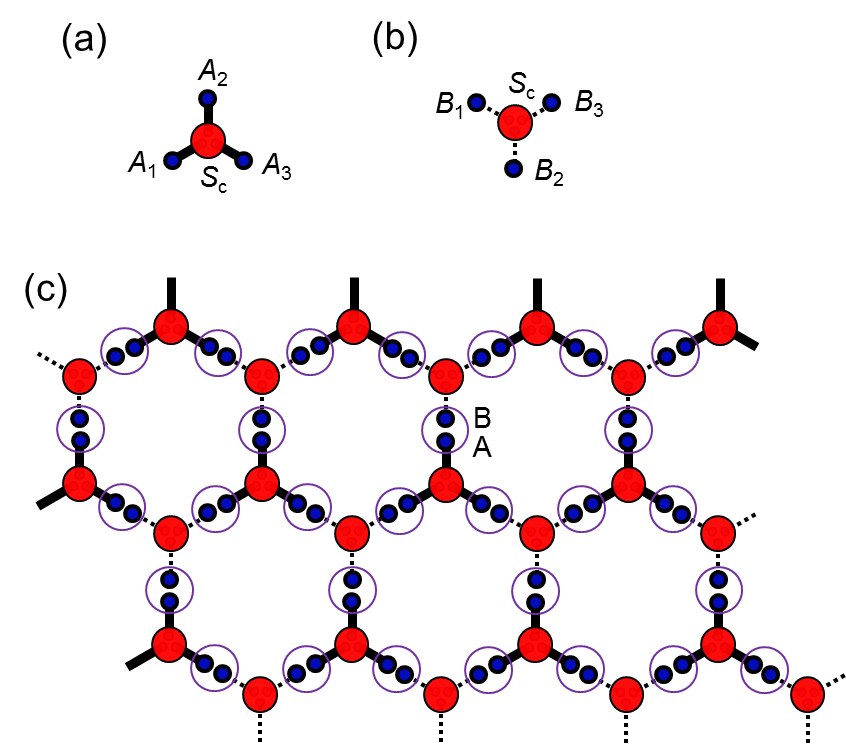}
  \caption{\label{fig:lattice1} Illustration of bottom-up approach. (a) \& (b) illustrate the building block of
  one unit, which consists
  of one center spin-3/2 and three  outer three virtual bond qubits. Two virtual bond qubits, each
  from a neighboring unit, form a physical bond spin-3/2 particle, as shown by circles enclosing them in (c).
  A two-dimensional or three-dimensional structure can be
  constructed. }
\end{figure}

 We have previously constructed a model Hamiltonian
  whose thermal states can be used for universal MBQC even without turning off the
Hamiltonian~\cite{Thermal}. The idea is to take a small unit of  a
few spins, e.g., one spin-3/2 $\vec{S}_i$ at the center coupled to
three outer spin-1/2 $\vec{s}_j$ that interact via the Heisenberg
interaction $\vec{S}_i\cdot \vec{s}_j$; see Fig.~\ref{fig:lattice1}.
Then we stack up many such units to form a higher dimensional
structure, e.g., the decorated 2D honeycomb or other trivalent
lattices,  or even 3D lattices, and then ``glue'' or map two smaller
spins (i.e. spin-1/2 particles) from neighboring units to single
larger spin; see e.g. Fig.~\ref{fig:lattice1}. Each merged spin,
which we shall refer to as a bond particle, possesses a Hilbert
space of dimension $4$ (i.e. two copies of a qubit) and hence is
equivalent to a spin-3/2 entity. One advantage of this approach is
that the
 ground state and its spectral gap can be readily
solved and checked. As we shall see, the exactly solvable
Hamiltonians thereby constructed allow for fault-tolerant, universal
quantum computation with thermal states, and even with topological
protection in three dimensions~\cite{Thermal}.

There was no free parameter
 in the Hamiltonians in Ref.~\cite{Thermal}. It was not clear whether or not such
 quantum computational universality only occurred for the specific Hamiltonian or could
 be extended to a region in a phase diagram. Here we use as building blocks
  two different types of interactions beyond the Heisenberg interaction to allow a free system parameter:
  the XXZ interaction
 $S^x_i s^x_j+S^y_i s^y_j+\Delta S^z_i s^z_j$ and an additional on-site anisotropic term $\vec{S}_i\cdot
\vec{s}_j -d_z (S_i^z)^2$, and investigate relation between the
statistical mechanical and quantum computational features of the
resultant two- and three-dimensional models as the system parameter
and the temperature vary. (Note the
 upper case $S_i$ is a spin operator for the center particle of larger spin
 magnitude, where $s_j$ is a spin-1/2 operator, i.e., `half' of the degree of freedom in a bond particle and
 will be denoted by $A$ or $B$ later).
 These interactions might be engineered in cold atoms or
 trapped ions.) It turns out to be useful to relate the ground state
 wavefunctions of the two models if we parameterize $\Delta$ by
 $\Delta=1+\delta$ in the first model and  thus the Heisenberg point is at $\delta=d_z=0$.

We thus arrive at two spin models. The Hamiltonian for model I
consists of two types of interactions: $H_{\rm I}=\sum_{{\rm
line}}V_{{\rm line}}+\sum_{{\rm dash}}V_{{\rm dash}}$, where
\begin{eqnarray}
V_{{\rm line}}&=& S_c^x A_b^x+S^y_c A_b^y+(1+\delta) S^z_c A_b^z\\
V_{{\rm dash}}&=& S_c^x B_b^x+S^y_c B_b^y+(1+\delta) S^z_c B_b^z,
\end{eqnarray}
where $A_b^\alpha$'s and $B_b^\beta$'s are two independent spin-1/2
operators for the two virtual qubits of a bond particle. For model
II, $H_{\rm II}= \sum_{{\rm line}}V_{{\rm line}}+\sum_{{\rm
dash}}V_{{\rm dash}}+\sum_{c} V_{c}$,
\begin{eqnarray}
V_{{\rm line}}&=& S_c^x A_b^x+S^y_c A_b^y+S^z_c A_b^z\\
V_{{\rm dash}}&=& S_c^x B_b^x+S^y_c B_b^y+S^z_c B_b^z\\
V_{c}&=& -d_z (S_c^z)^2,
\end{eqnarray}
the $V_c$ is a local term on the center particles. These two models
can be placed on two- and three-dimensional lattices; see e.g. the
hexagonal lattice in Fig.~\ref{fig:lattice1} and the 3D lattice in
Fig.~\ref{fig:lattice2}c.
\subsection{Model 1: XXZ interaction in a building block}
Consider the XXZ
 interaction for each unit.  The Hamiltonian within each unit can be exactly solved.
 For $\delta >-2$ the ground state energy  is
$E_0(\delta)=(-9-5\delta -2 \sqrt{9 +4 \delta^2})/4$ (see
Fig.~\ref{fig:E}) and the ground-state wavefunction for a unit
(which is unique and gapped) is
\begin{eqnarray}
\label{eqn:PsiDelta} && |\Psi(\delta)\rangle=N_0(\delta)\Big[
-(\ket{3/2,-3/2}-\ket{-3/2,3/2}) \nonumber\\
&& +\frac{-2\delta +\sqrt{9 +4
\delta^2}}{3}(\ket{1/2,-1/2}-\ket{-1/2,1/2})\Big],
\end{eqnarray}
where $N_0(\delta)$ is a normalization constant such that the
wavefunction is properly normalized, the symbol $\ket{m_c,m_s}$
denotes the joint state of the center spin-3/2 ($\ket{m_c}$) and
three outer virtual spin-1/2 particles (collectively denoted by
$\ket{m_s}$). The examples of the latter are,
$\ket{m_s=3/2}=|\uparrow\uparrow\uparrow\rangle$ and
$\ket{m_s=1/2}=(\ket{\uparrow\uparrow\downarrow}+\ket{\uparrow\downarrow\uparrow}+\ket{\downarrow\uparrow\uparrow})/\sqrt{3}$.
The ground-state wavefunction for the whole 2D system is simply a
product of $|\Psi(\delta)\rangle$ over all units (modulo appropriate
merging).

For $\delta\gg 1$, $|\Psi(\delta)\rangle\sim |3/2,-3/2\rangle -
|-3/2,3/2\rangle$, which is a four-spin GHZ state. Because of the
merging of outer spin-1/2 particles across two units, such
entanglement is useful for quantum computation, as explained in
Refs.~\cite{GoChua} and~\cite{VerstraeteCirac}. As $\delta$
approaches $0$, it reduces to Heisenberg interaction within a unit
and universal quantum computation can be done on such a
two-dimensional structure~\cite{Thermal}.

 For $\delta <-2$,
the ground states are doubly degenerate:
$\ket{3/2,3/2}=|\!\Uparrow\uparrow\uparrow\uparrow\rangle$ and
$\ket{-3/2,-3/2}=|\!\Downarrow\downarrow\downarrow\downarrow\rangle$,
each of which is ferromagnetic within the unit (where we have used
$\Uparrow$ and $\Downarrow$ to denote the $|\pm3/2\rangle$ of the
center particle). The ground-state energy is $E_0=9(1+\delta)/4$. At
a small but finite temperature $T$ (smaller than the gap above the
ground space), the thermal state will be approximately
$1/2\ketbra{\Uparrow\uparrow\uparrow\uparrow}+
1/2\ketbra{\Downarrow\downarrow\downarrow\downarrow}$, possessing no
entanglement. Therefore, for $\delta<-2$, the whole system is not
useful for universal quantum computation due to {\it lack} of
entanglement.

The ground state energy has discontinuity in its first-order
derivative with respect to $\delta$ (see Fig.~\ref{fig:E}), i.e.,
there is a first-order phase transition. As the ground state in the
ferromagnet-like phase, $\delta <-2$, cannot enable universal
quantum computation, one is led to inquire whether universal quantum
computation is possible for $\delta>-2$ and whether emergence of
such computational power coincide with the phase transition.

\subsection{Model 2: Heisenberg interaction with an on-site anisotropic term}
\begin{figure}
   \includegraphics[height=5cm]{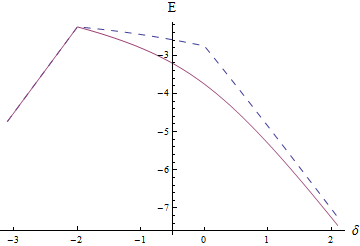}
    \includegraphics[height=5cm]{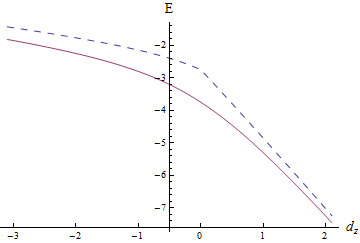}
  \caption{\label{fig:E} Ground-state and the first excited-state energies in each unit for the two
  models (top: model 1; bottom: model 2). The difference in the two energies is also the gap of the
  corresponding two- or three-dimensional models (for any finite system as well as in the thermodynamic limit).
  For the first model, the lowest two energy levels are degenerate for $\delta\le-2$. The ground-state energy
  exhibits discontinuity in its first derivative w.r.t. $\delta$,
  implying a first-order quantum phase transition in the model.
  However, for the second model, the ground-state energy is
  analytic for all range of $d_z$, implying non-existence of  phase transition.
   }
\end{figure}

In this section we consider interaction of the form $\vec{S}_i\cdot
\vec{s}_j -d_z (S_i^z)^2$. As Pauli operators square to identity
$\sigma_\mu^2=\openone$, there is no need to add a term $(s_i^z)^2$
for spin-1/2 particles. As it is exactly solvable, the ground state
energy for a unit consof one center spin-3/2 and three outer
spin-1/2 particles is $E_0(d_z)=(-9-5d_z -2 \sqrt{9+4 d_z^2})/4$ for
all range of $d_z$; see Fig.~\ref{fig:E}. Furthermore, the ground
state (which is unique and gapped) is
\begin{eqnarray}
&&|\Psi(d_z)\rangle=N_1(d_z)\Big[ -(\ket{3/2,-3/2}-\ket{-3/2,3/2})\nonumber\\
&&+\frac{-2 d_z +\sqrt{9 +4
d_z^2}}{3}(\ket{1/2,-1/2}-\ket{-1/2,1/2})\Big],
\end{eqnarray}
where $N_1(d_z)$ is a normalization constant such that the
wavefunction is properly normalized. We see that the ground state
wavefunction and its energy are of the same form as in model 1 when
$\delta>-2$. Hence, the computational power of the two models at
zero temperature will be the same in the corresponding range.
However, in contrast with model 1, this model does not have a phase
transition in the state of matter. As this model contains the
Heisenberg point, which is universal for MBQC, one is led to inquire
whether the whole phase is universal (as there is no phase
transition), as opposed to the first model.

\section{Creating a 2D cluster state from ground states}
\label{sec:ground} We shall first consider the range of the
parameters for the two models where ground state is of the same form
within a unit:
\begin{eqnarray}
 |\Psi(a)\rangle\sim&&
-(\ket{3/2,-3/2}-\ket{-3/2,3/2})\nonumber \\
&&+\frac{1} {a}(\ket{1/2,-1/2}-\ket{-1/2,1/2}).\label{eqn:wfa}
\end{eqnarray}
For Model 1: the relation of $a$ to $\delta$ (for $\delta>-2$) is
given by
\begin{equation}
a^{-1}=\frac{-2\delta +\sqrt{9 +4 \delta^2}}{3}.
\end{equation}
For Model 2: the relation of $a$ to $d_z$ (for all range of $d_z$)
is given by
\begin{equation}
a^{-1}=\frac{-2 d_z +\sqrt{9 +4 d_z^2}}{3}.
\end{equation}
Since the two models possess the same form of the ground-state
wavefunction in the appropriate range of the parameters, we can deal
with the quantum computational universality at zero temperature with
equal footing. We note that, however, at finite temperatures the
region of quantum computational universality will differ due to the
different structures in the excited states and their energies. This
will be treated in the next section.

The case $\delta=d_z=0$ reduces to the Heisenberg interaction and
the use for MBQC has been shown and detailed in Ref.~\cite{Thermal}
and this corresponds to $a=1$. Examining the
wavefunction~(\ref{eqn:wfa}), we see that we can recover the $a=1$
wavefunction if we can apply the following operation on the center
spin-3/2 particle:
\begin{equation}
D(a)={\rm diag}(1,a,a,1),
\end{equation}
in the basis of $\ket{3/2}$, $\ket{1/2}$, $\ket{-1/2}$, and
$\ket{-3/2}$. However, such a filtering operation cannot be realized
with unit probability of  success.  This is because to implement a
filtering operation such as $D(a)$, one needs to include another
 element $D'(a)$ to represent the unsuccessful filtering so that
$D(a)^\dagger D(a) + D'(a)^\dagger D'(a)=\openone$.

The solution is to use generalized measurement that can incorporate
the filtering. For $a=1$, the filtering is not needed and a
generalized measurement has been used~\cite{Thermal}  so that a GHZ
state, such as $(\ket{3/2,-3/2}-\ket{-3/2,3/2})$, can be obtained
within each unit. The POVM elements $\tilde{F}_\alpha$ (for
spin-3/2's) were first constructed in
Refs.~\cite{WeiAffleckRaussendorf11,Miyake11},
\begin{subequations}
\label{POVM2}
  \begin{eqnarray}
\tilde{F}_{x}&=&\sqrt{\frac{2}{3}}(\ket{3/2}_x\bra{3/2}+\ket{-3/2}_x\bra{-3/2}) \\
\tilde{F}_{y}&=&\sqrt{\frac{2}{3}}(\ket{3/2}_y\bra{3/2}+\ket{-3/2}_y\bra{-3/2})\\
\tilde{F}_{z}&=&\sqrt{\frac{2}{3}}(\ket{3/2}_z\bra{3/2}+\ket{-3/2}_z\bra{-3/2}).
\end{eqnarray}
\end{subequations}
For general $a$, we use a deformed POVM with elements
$F_\alpha=q_\alpha(a) \tilde{F}_\alpha D(a)$ ($\alpha=x,y,z$ and the
proportional constants $q_\alpha(a)$ are to be determined below) to
act on the center particle so as to distill a GHZ state. The reason
that $\tilde{F}_\alpha D(a)$ works can be illustrated by the example
$\alpha=z$. First $D(a)$ restores the wavefunction back to the $a=1$
case. Then $\tilde{F}_z$ filters out the GHZ state
$(\ket{3/2,-3/2}-\ket{-3/2,3/2})$, or equivalently,
$\ket{3/2}\ket{\downarrow\downarrow\downarrow}-\ket{-3/2}\ket{\uparrow\uparrow\uparrow}$.
If we choose to encode the effective qubit for the center particle
by $\ket{0}\equiv -\ket{-3/2}$ and $\ket{1}\equiv \ket{3/2}$, and
for the virtual spin-1/2's by the usual definition
$|0\rangle\equiv\ket{\uparrow}$ and
$|1\rangle\equiv\ket{\downarrow}$, then the resultant GHZ state for
$F_z$ outcome is
\begin{equation}\label{eqn:GHZ}
|{\rm GHZ}\rangle=\frac{1}{\sqrt{2}}(|0000\rangle+|1111\rangle).
\end{equation}
As the $a=1$ wavefunction is symmetric under rotation, the case of
$\alpha=x,y$ simply produces the GHZ state in the $x$ and $y$ bases,
respectively. By imposing the completeness relation, $\sum_\alpha
F_\alpha^\dagger F_\alpha =\openone$, we find $q_x(a)=q_y(a)=1/a$
and $q_z(a)=\sqrt{(3a^2-1)/(2a^2)}$. This can be verified easily by
direction calculation that yields
\begin{eqnarray}
&& F_x^\dagger F_x +F_y^\dagger F_y=
\begin{pmatrix}
\frac{1}{3a^2} & 0 & 0 & 0\cr
 0 & 1 & 0 & 0\cr
 0& 0 &  1 & 0\cr
 0& 0 & 0 & \frac{1}{3a^2}
\end{pmatrix}\\
&& F_z^\dagger F_z =
\begin{pmatrix}
1-\frac{1}{3a^2} & 0 & 0 & 0\cr
 0 & 0 & 0 & 0\cr
 0& 0 & 0 & 0\cr
 0& 0 & 0 & 1-\frac{1}{3a^2}
\end{pmatrix}.
\end{eqnarray}
In order for the above expressions to remain non-negative, such
construction is valid only when $3a^2\ge 1$. We note that a similar
construction of POVM has been first used in Ref.~\cite{Darmawan} in
the context of a deformed AKLT model.
 The  POVM $F_\alpha$'s give rise to three possible outcomes, and any of
them is a good outcome. This effectively generates product of GHZ
states among all units. However, the GHZ states are in different
bases, depending on the outcome $\alpha$. To fix the outcome basis
in the $z$ basis, we can imagine applying a unitary transformation
to the post-POVM state if $\alpha=x$ or $y$. This is equivalent to
using a different measurement basis for the effective qubits. For
outcomes $F_x$ and $F_y$, we perform operations $U_y$ and $U_x$,
respectively, where
\begin{eqnarray}
U_y &=& \exp [i\frac{\pi}{2}(S^y_c+s^y_1+s^y_2+s^y_3)]\\
U_x &=& \exp [-i\frac{\pi}{2}(S^x_c+s^x_1+s^x_2+s^x_3)],
\end{eqnarray}
so that the resultant GHZ state is always of the
form~(\ref{eqn:GHZ}). Then measurement on the bond particle (i.e., a
joint measurement on the two virtual qubits) can be used to induce a
control-Z (CZ) gate between two center
particles~\cite{Cai10,Thermal,WeiRaussendorfKwek11}. The result is a
cluster state, a universal resource state. To summarize, we have
thus shown that for $a\ge 1/\sqrt{3}$, the ground state is universal
for MBQC.

For $a<1/\sqrt{3}$, we need to use a different POVM~\cite{Darmawan},
i.e.,
\begin{eqnarray}
&& F'_x=\sqrt{3}\,\tilde{F}_x D(a), \ F'_y=\sqrt{3}\, \tilde{F}_y D(a)\\
 && F'_z =
\begin{pmatrix}
0 & 0 & 0 & 0\cr
 0 & \sqrt{1-3a^2} & 0 & 0\cr
0& 0 &\sqrt{1-3a^2} & 0\cr
 0& 0 & 0 &0
\end{pmatrix}.
\end{eqnarray}
One can easily verify the completeness relation $\sum_\alpha
{F'}_\alpha^\dagger {F'}_\alpha=\openone$, via a direct calculation
that yields
\begin{eqnarray}
&& {F'}_x^\dagger F'_x +{F'}_y^\dagger F'_y=
\begin{pmatrix}
1& 0 & 0 & 0\cr
 0 & 3a^2 & 0 & 0\cr
 0& 0 &  3a^2 & 0\cr
 0& 0 & 0 & 1
\end{pmatrix}\\
&& {F'}_z^\dagger {F'}_z =
\begin{pmatrix}
0 & 0 & 0 & 0\cr
 0 & 1-3a^2 & 0 & 0\cr
 0& 0 & 1-3a^2 & 0\cr
 0& 0 & 0 & 0
\end{pmatrix}
\end{eqnarray}
 In this case, the outcome $F'_z$ is not
desirable, i.e., it needs to be regarded as an error (specifically a
qubit loss) as a GHZ cannot be obtained. The outcomes from $F'_x$
and $F'_y$ still yield a perfect GHZ state. To arrive at the same
GHZ state~(\ref{eqn:GHZ}) as in the case of $a\le 1/\sqrt{3}$, we
further perform operations $U_y$ and $U_x$, for outcomes $F_x'$ and
$F_y'$, respectively. Sites with undesirable outcome $F'_z$ is
equivalent to having leakage out of logical qubit space (or a qubit
loss) but can be removed without affecting neighboring center sites
by performing measurement on the surrounding bond particles so as to
disentangle the unit (the center spin and the three virtual qubits)
from the neighboring ones. Thus the qubit loss rate corresponds to
the probability $p_{\rm delete}$ of obtaining a $F_z'$ outcome,
where
\begin{equation}
p_{\rm delete}=\frac{1- 3 a^2}{1+a^2}.
\end{equation}
 If $1- p_{\rm delete}$ is smaller than the site percolation
threshold $p_{\rm th}^{(\rm site)}$ (which depends on the lattices,
such as honeycomb, cross, and square-octagon), then there is not
sufficient {\it connection\/} in the remaining network and thus no
two-dimensional graph state can be distilled~\cite{Browne}.
Fortunately, it turns out that there is a finite range of $a$ below
$1/\sqrt{3}$ such that the remaining sites still possess enough
connection, i.e., the corresponding graph resides in the
supercritical phase of percolation. For universal MBQC, it is thus
required that $p_{\rm delete} \le 1- p_{\rm th}^{(\rm site)}$, i.e.,
$a^2\ge p_{\rm th}^{(\rm site)}/ (4- p_{\rm th}^{(\rm site)})$. This
gives $a^2\gtrsim 0.211, 0.223, 0.229$ for honeycomb,
square-octagon, and cross lattices, respectively. For the honeycomb
lattice, the threshold translates to  $\delta=d_z\gtrsim -1.2882$.
Therefore, at zero temperature, there is a transition in the quantum
computational power in both models and, in the first model, before
the system reaches its phase transition at $\delta=-2$ as $\delta$
decreases. The exact location of the transition point in the quantum
computational power depends on the underlying lattices, due to the
connection to percolation. Connection to percolation and quantum
computation has been previously explored, such as site percolation
in noisy cluster states~\cite{Browne} and bond percolation in
nondeterministic gates for cluster state preparation~\cite{Kieling}.

The above analysis shows that for model 1 the transition in the
quantum computational power (at $\delta\approx -1.2882$ on the
honeycomb lattice) does not coincide with the transition in the
phase of matter (at $\delta=-2$). Moreover,  even though model 2
possesses only one phase of matter, the quantum computational
universality only exists in part of the phase. In the following we
shall investigate the finite-temperature effect on the quantum
computational universality and determine the corresponding `phase
diagram' in terms of quantum computational capability.

\section{Thermal states and fault tolerance: two and three dimensions}
\label{sec:Thermal} 
\begin{figure}
\vspace{1cm}
   \includegraphics[width=8cm]{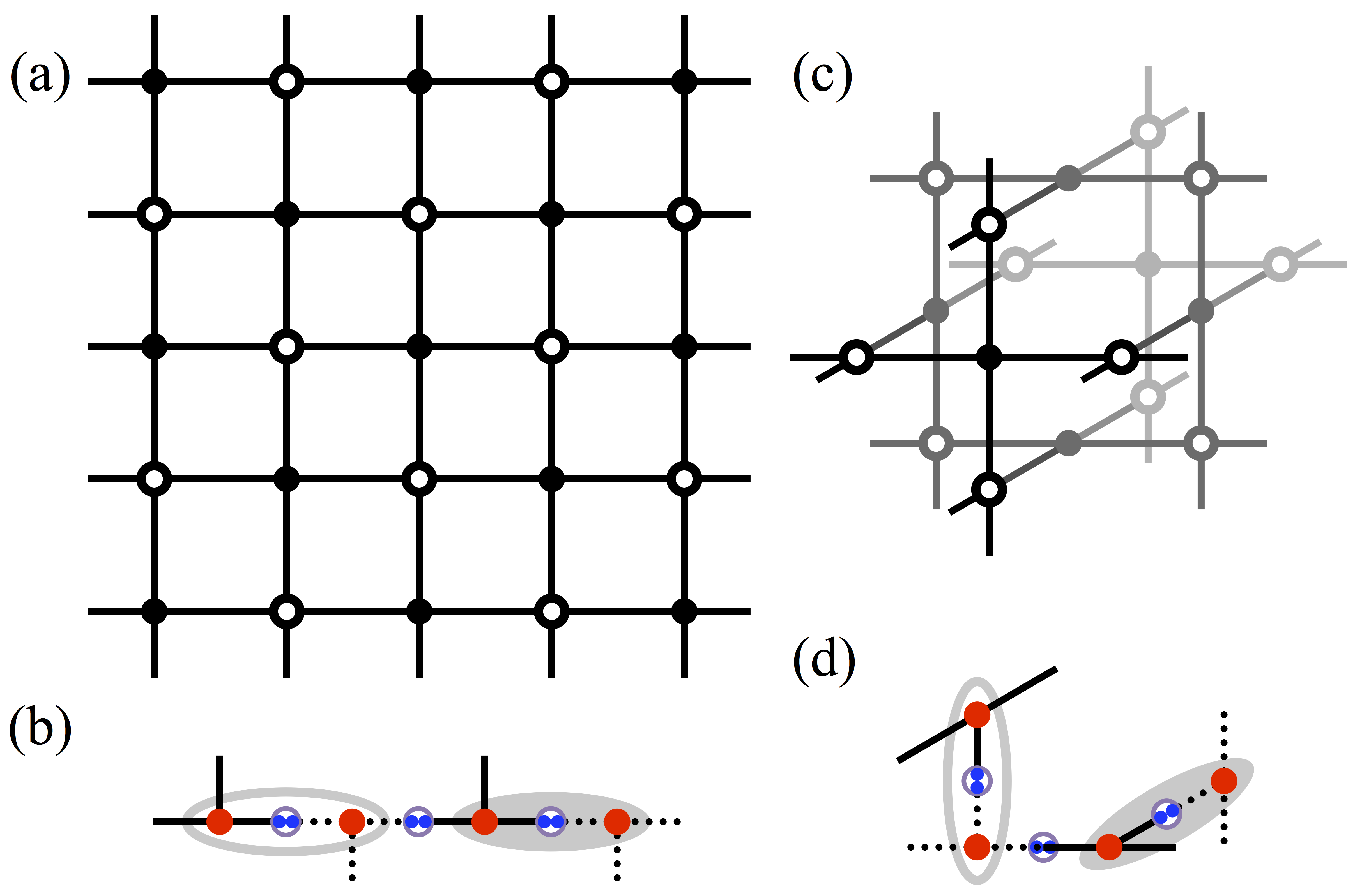}
  \caption{Lattices of cluster states and microscopic construction of qubits.  (a) is the 2D square-lattice
  cluster state,
  and (c) is the unit cell of the 3D bcc (body-center cubic lattice) cluster state. Both could be constructed
  from trivalent lattices (with degree-3 nodes) using the modification as shown in (b) and (d), respectively, which converts
  a degree-4 node to combinations of degree-3 nodes; see also Ref.~\cite{Fujii}. Note that with
  such a construction, (a) will be turned to a brickwall structure
  equivalent to the honeycomb lattice shown in
  Fig.~\ref{fig:lattice1}c. See also Fig.~\ref{fig:honeysquare} and
  the Appendix for the conversion.
  {In the above (b) and (d)}, each logical qubit (oval) is composed of spins from two units. The spins inside
  each oval will eventually be converted to one logical qubit. The
  circle (which includes two virtual qubits, i.e., a single bond particle) between the two ovals
  are used to
  entangle neighboring two effective qubits (within the ovals) via measurement of a bond particle.
  Essentially, the product of two
  GHZ states from two such units can be converted to a single GHZ via local measurement on the bond particle inside the oval as well as one of the center
  particle. }
  \label{fig:lattice2}
\end{figure}

\begin{figure}
\vspace{1cm}
   \includegraphics[width=0.45\textwidth]{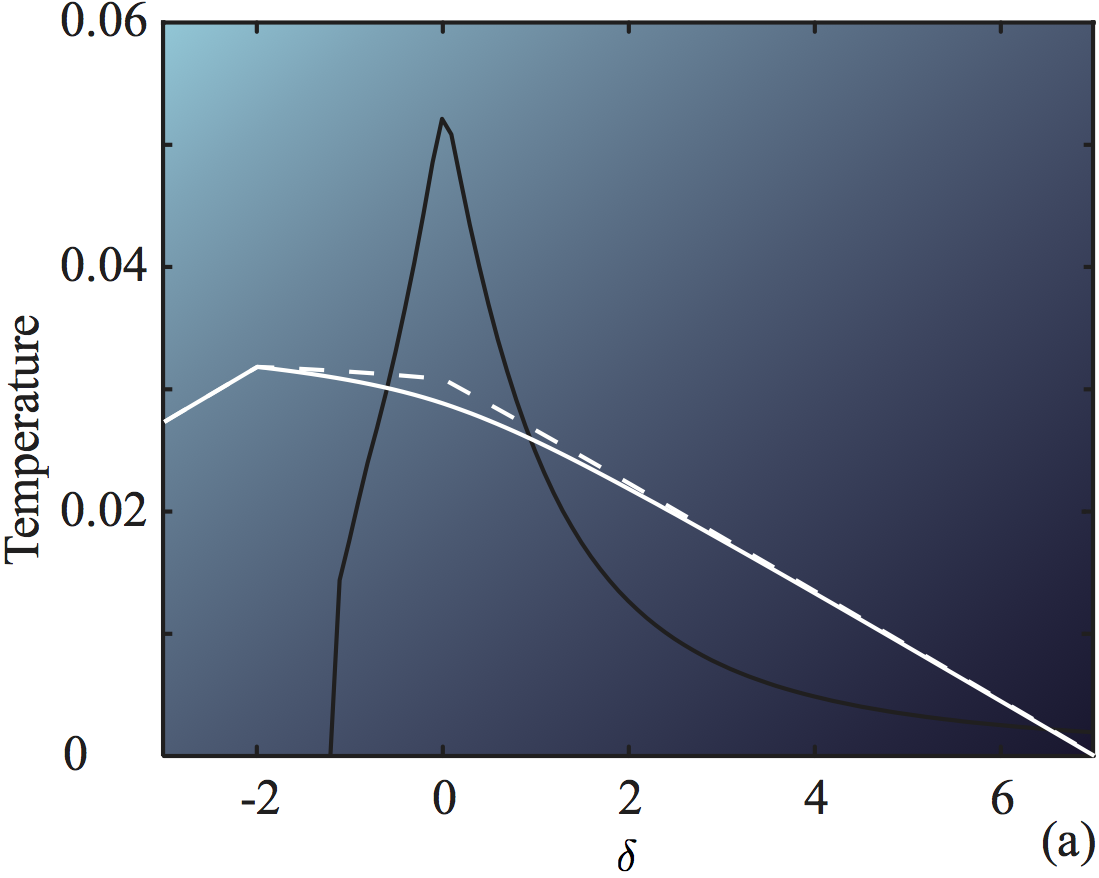}\\
   \includegraphics[width=0.45\textwidth]{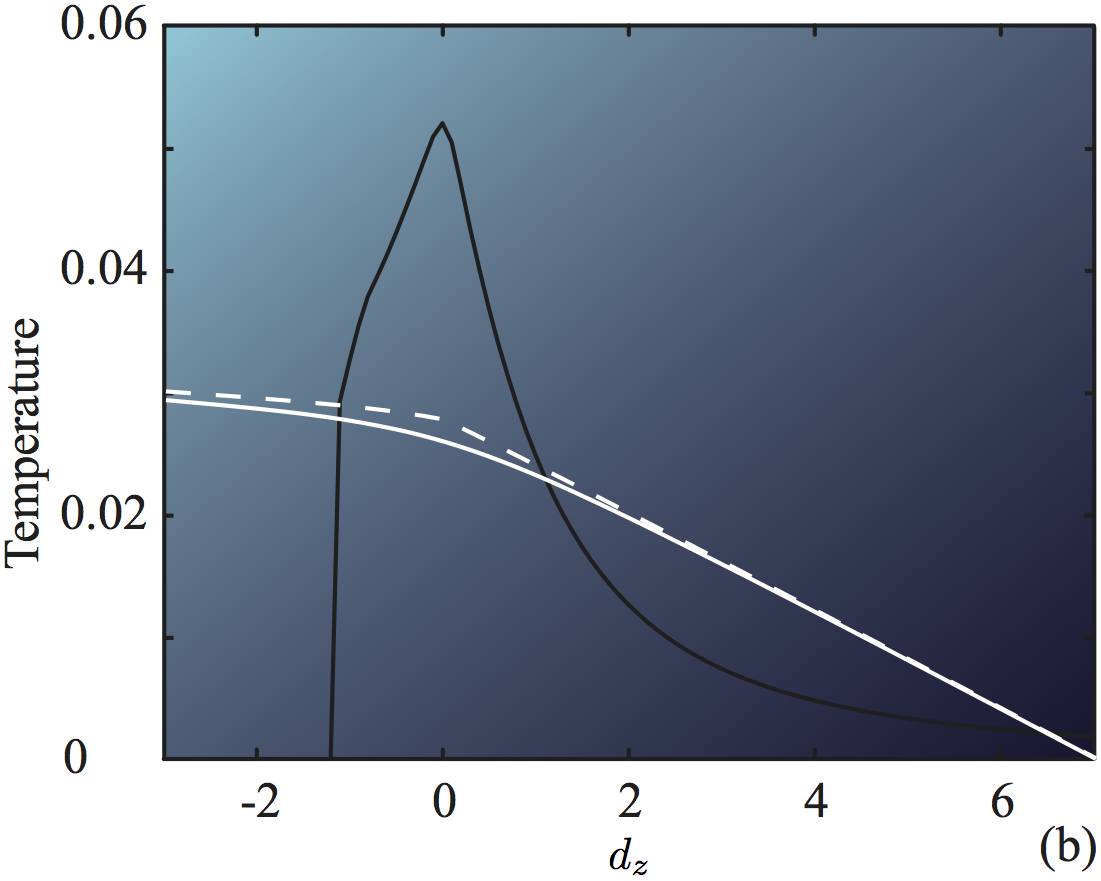}
  \caption{\label{fig:2Dphase} 2D Phase diagrams in terms of computational
  power:  (a) Upper: the XXZ model; (b) bottom: the anisotropic model. In the region below the solid black curve the equilibrium thermal
  states provide universal resource for MBQC. For reference, the
 energies for the ground state and the first excited is shown in
 white curves (solid and dashed, respective).
    }
\end{figure}

\begin{figure}
\vspace{1cm}
   \includegraphics[width=0.45\textwidth]{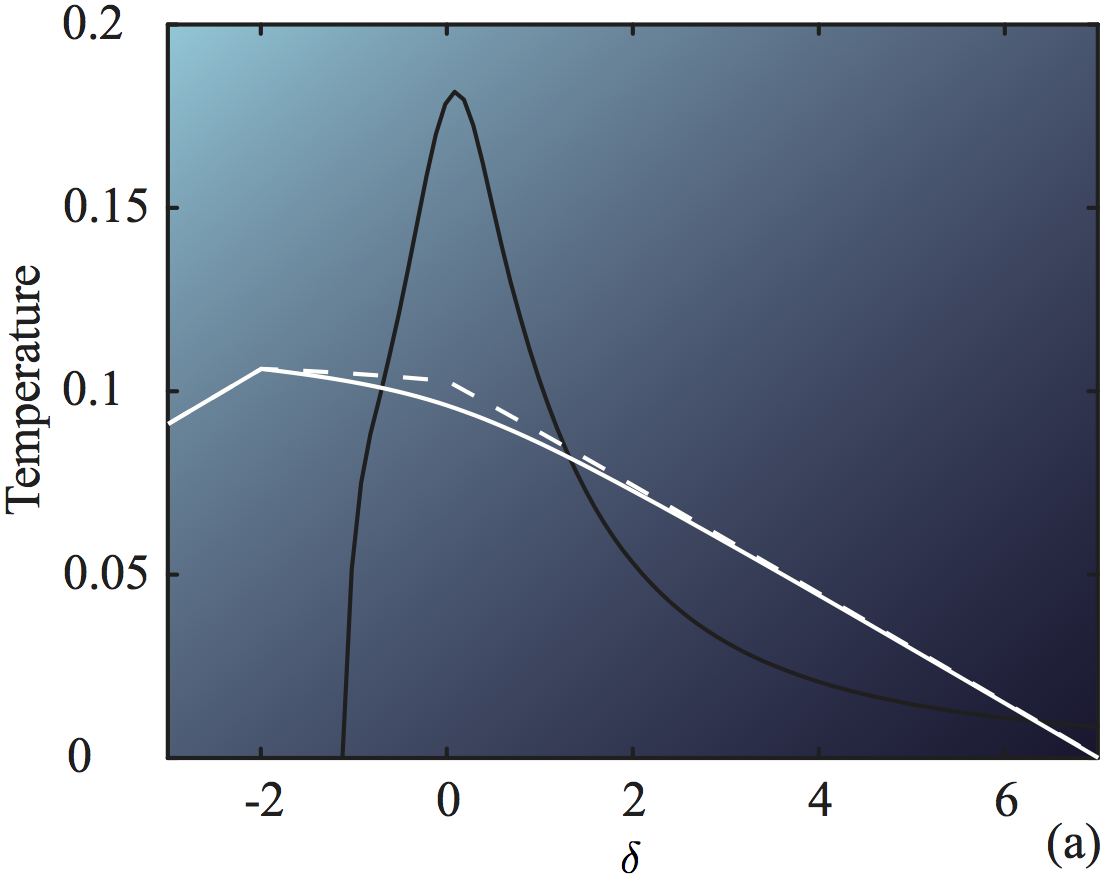}
   \includegraphics[width=0.45\textwidth]{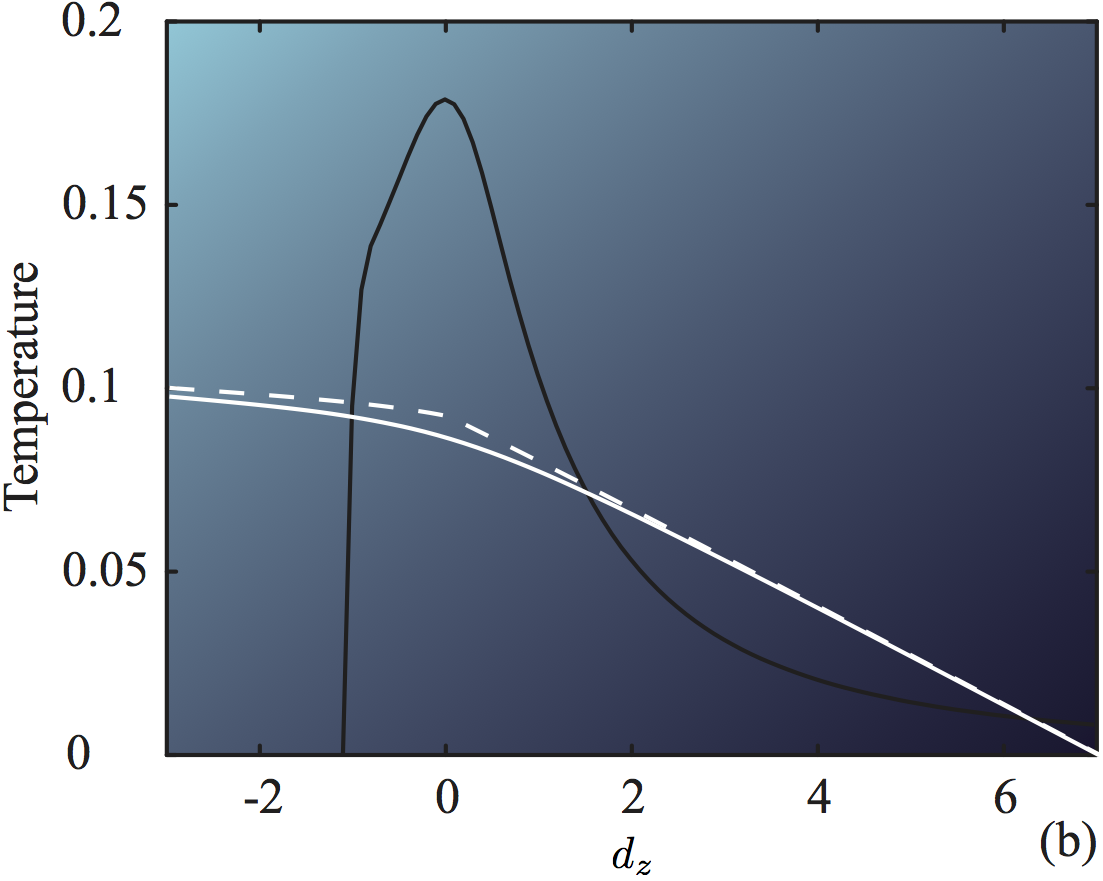}
  \caption{\label{fig:3Dphase} 3D Phase diagrams in terms of computational power: (a) Upper: the XXZ model; (b) bottom: the anisotropic model.
  (a) Upper: the XXZ model; (b) bottom: the anisotropic model. In the region below the solid black curve the equilibrium thermal
  states provide universal resource for MBQC. For reference, the
 energies for the ground state and the first excited is shown in
 white curves (solid and dashed, respective). Note that the transition temperatures are about 3 to 4 times higher than in 2D at the Heisenberg point. }
\end{figure}

Because of the structure that the Hamiltonian
for both models can be divided into units independent with one
another, the free energy at finite temperatures are non-singular,
and thus there are no phase transition at finite temperatures. As we
shall see below, the region with universal quantum computational
power exists up to certain finite temperatures. For a finite
temperature, the system is not in the exact ground state but a
thermal state. This means that the production of a GHZ state in each
unit (and thus the global cluster state) is faulty. Therefore,
whether the `phase' of universal quantum computational power exists
depends on how one can deal with errors. In particular, the `phase'
boundary will depend on the error rates and the thresholds for
fault-tolerant quantum computation (FTQC). Our goal is to establish
the existence of a nonzero-temperature region that universal MBQC is
possible rather than to pin point the absolute boundary of such a
region. In the following we describe in detail the error analysis
and how the `phase diagram' of the computational power is obtained.
For those readers who wish to skip the details, the `microscopic'
construction is in Fig.~\ref{fig:lattice2} and the resultant phase
diagrams for both models are in Fig.~\ref{fig:2Dphase} for 2D and in
Fig.~\ref{fig:3Dphase} for 3D.

For each set of particles, the thermal state reads
\begin{equation}
\rho_T = \frac{e^{-H/T}}{\tr\, e^{-H/T}},
\end{equation}
where $H$ is the Hamiltonian of four spins including one spin-3/2
and three spin-1/2, and $T$ is the temperature. As the input state
is a thermal state, the output state after the POVM and the
associated unitary operations is a noisy GHZ state. If
$a>1/\sqrt{3}$, the output state is
\begin{equation}
\rho_{GHZ} = U_y F_x \rho_T F_x^{\dag} U_y^{\dag} + U_x F_y \rho_T
F_y^{\dag} U_x^{\dag} + F_z \rho_T F_z^{\dag},
\end{equation}
and the success probability is $p_s=1$.
If $a<1/\sqrt{3}$, the output state is
\begin{equation}
\rho_{GHZ} = p_s^{-1} (U_y F'_x \rho_T F_x^{\prime \dag} U_y^{\dag}
+ U_x F'_y \rho_T F_y^{\prime \dag} U_x^{\dag}),
\end{equation}
where the success probability is $p_s=1-\tr F'_z \rho_T F_z^{\prime
\dag}$, due to `loss' of logical quits.
\begin {table}
\begin{center}
\begin{tabular}{ | c | c || c | c | }
\hline $\sigma$ & Probability & $\sigma$ & Probability
\\ \hline
$\openone$ & $0.9942$ & $Z_0$ & $3.45\times 10^{-3}$
\\ \hline
$X_1$ & $3.84\times 10^{-4}$ & $Z_0X_1$ & $3.84\times 10^{-4}$
\\ \hline
$X_2$ & $3.84\times 10^{-4}$ & $Z_0X_2$ & $3.84\times 10^{-4}$
\\ \hline
$X_3$ & $3.84\times 10^{-4}$ & $Z_0X_3$ & $3.84\times 10^{-4}$
\\ \hline
$X_1X_2$ & $2.38\times 10^{-9}$ & $Z_0X_1X_2$ & $2.38\times 10^{-9}$
\\ \hline
$X_2X_3$ & $2.38\times 10^{-9}$ & $Z_0X_2X_3$ & $2.38\times 10^{-9}$
\\ \hline
$X_1X_3$ & $2.38\times 10^{-9}$ & $Z_0X_1X_3$ & $2.38\times 10^{-9}$
\\ \hline
$ X_0$ & $1.23\times 10^{-16}$ & $ Z_0X_0$ & $1.15\times 10^{-16}$
\\ \hline
\end{tabular}
\caption{ The Pauli operators $\{\sigma\}$ that appear in
Eq.~(\ref{eq:PE}). $X$ denotes Pauli $\sigma_x$ and $Z$ denotes
Pauli $\sigma_z$. There are in total $15$ different inequivalent
errors that may occur on the noisy GHZ state. Subscript $0$ denotes
the center spin (i.e. $c$ in Fig.~\ref{fig:lattice1}a), and
subscripts $1-3$ denote the surrounding virtual qubits. The list
only considers inequivalent errors; e.g., $Z_0$ and $Z_{i=1,2,3}$
errors produce the same consequence for the GHZ
state~(\ref{eqn:GHZ}), and hence either one of them, say, $Z_0$, is
needed in Eq.~(\ref{eq:PE}) and listed in the table.  Moreover, the
triple X error $X_1X_2X_3$ is equivalent to a single $X_0$ error,
i.e., $X_0\equiv X_1X_2X_3$. The probabilities of these inequivalent
errors for the example of the isotropic model, $\delta=d_z=0$, are
also listed at a temperature $T=0.16$. }
\end{center}
\label{tab:list}
\end {table}

The ideal GHZ state $\ket{GHZ}$~(\ref{eqn:GHZ}) is the common
eigenstate the  stabilizer elements $X_0X_1X_2X_3$, $Z_0Z_1$,
$Z_0Z_2$, and $Z_0Z_3$ (this set denoted by $\{K\}$) with the same
eigenvalue $+1$. Here, $X_0$ and $Z_0$ are Pauli operators of the
center qubit, and similarly for other three qubits. In order to use
the fault-tolerant quantum computing (FTQC) theory to analyze the
computational power, we convert imperfections in the noisy GHZ state
$\rho_{GHZ}$ into Pauli errors by randomly performing stabilizer
operations, which results in
\begin{equation}
\rho^{\prime}_{GHZ} = \prod_{K\in\{K\}} \frac{1}{2}([\openone] +
[K]) \rho_{GHZ},
\end{equation}
where $[\Lambda]\rho \equiv \Lambda\rho\Lambda^\dag$. Here, $\{K\}$
is the set of the above stabilizer generators. Such randomization
can be effectively performed by updating the basis of the ensuing
single-particle measurements rather than actively by actively
applying $K$'s. The state $\rho^{\prime}_{GHZ}$ is thus diagonal in
the basis of stabilizers and can be written as
\begin{equation}
\rho^{\prime}_{GHZ} =
\sum_{\sigma\in\{\sigma\}}p_{\sigma}[\sigma]\ketbra{GHZ},
\label{eq:PE}
\end{equation}
where $\{\sigma\}$ are Pauli operators listed in Table
\ref{tab:list}, each $[\sigma]\ketbra{GHZ}$ corresponds to a common
eigenstate of stabilizers, and $p_{\sigma\neq \openone}$ is the
probability of the corresponding Pauli error. If the eigenvalue of
$X_0X_1X_2X_3$  is $-1$ in an eigenstate, there is an error $[Z_0]$
in the state. Note that for convenience of notation we use $X$ to
denote the Pauli $\sigma_x$ and $Z$ the Pauli $\sigma_z$ and one
could also attribute $-1$ eigenvalue of $X_0X_1X_2X_3$ to an
$Z_{i=1,2,3}$ error instead of $Z_0$, but it is equivalent.
Similarly, the $-1$ eigenvalue of $Z_0Z_i$ corresponds to an $[X_i]$
error. Therefore, error probabilities can be obtained from diagonal
elements of $\rho^{\prime}_{GHZ}$.

{As seen above, in addition to single-qubit errors, some errors
occur simultaneously, such as $[Z_0X_i]$ and $[X_iX_j]$. In our
numerical results, we find that only the $[Z_0X_i]$ type of
correlated errors are significant (see e.g. Table~\ref{tab:list}),
and other correlated errors are negligible even at the transition
point of the computation power, i.e., the FTQC threshold. Actually,
these other correlated errors constitute less than $3\%$ of the
overall errors. Therefore, only the errors $[Z_0]$, $[X_i]$, and
$[Z_0X_i]$ will be taken into account in the following.}

We can construct a 2D cluster state on the square lattice from the
models sitting on the honeycomb lattice, as well as 3D cluster state
from the models on the lattice proposed in Ref.~\cite{Fujii},
modified from a construction in Ref.~\cite{Thermal} (see Fig.
\ref{fig:lattice2}). In both cases, each qubit on the cluster state
corresponds to two GHZ states. The procedure for obtaining a 2D
cluster state on a square lattice is explained in detail in the
Appendix, and is easily adapted to the 3D case. Moreover, the effect
of errors on the cluster state can be analyzed straightforwardly,
and is summarized in Table~\ref{tbl:error}. We describe them now.
{On the two GHZ states, each $[Z_0]$ error is propagated to a $[Z]$
error on the corresponding cluster-state qubit. We label the
spin-1/2 bond particle measured for fusing two GHZ states as
qubit-1. Then, each $[X_1]$ error is propagated to a correlated
error $[ZZ]$ on two neighbouring cluster-state qubits (see Table
\ref{tbl:error}). And, $[X_2]$ and $[X_3]$ errors are propagated to
independent $[Z]$ errors on neighbouring cluster-state qubits.
Similarly, each $[Z_0X_1]$ is propagated to a correlated error
$[ZZZ]$ on the corresponding cluster-state qubit and two neighboring
cluster-state qubits, and $[Z_0X_2]$ and $[Z_0X_3]$ errors are
propagated to a correlated error $[ZZ]$ on the corresponding
cluster-state qubit and one neighboring cluster-state qubit. Other
types of errors on GHZ states have been neglected as they rarely
occur. Therefore, on the final cluster state, the total probability
of phase errors on each qubit is
\begin{eqnarray}
p_z &\simeq & 2( p_{Z_0} + 2p_{X_1} + p_{X_2} + p_{X_3} \notag \\
&& + 3p_{Z_0X_1} + 2p_{Z_0X_2} + 2p_{Z_0X_3} ),
\end{eqnarray}
where $p_{Z_0}$, $p_{X_i}$, and $p_{Z_0X_i}$ are probabilities of
errors $[Z_0]$, $[X_i]$, and $[Z_0X_i]$ on each GHZ state,
respectively. The overall factor of 2 comes from the usage of two
units to build one qubit in the cluster state. On the finial cluster
state, there exist (i) correlated errors $[ZZ]$ with a probability
$2p_{X_1}$ on some pairs of qubits connected to the same qubit, (ii)
correlated errors $[ZZ]$ with a probability $2p_{Z_0X_2}$ or
$2p_{Z_0X_3}$ on each pair of directly connected qubits, and (iii)
correlated errors $[ZZZ]$ with a probability $2p_{X_1}$ on some
trimers formed by connected qubits. All the contribution of
correlated errors to each single qubit has been included in $p_z$.}
Furthermore, because a cluster-state qubit is missing if one or two
GHZ states are not successfully generated, the loss rate of cluster
qubits is $p_l = 1-p_s^2$.

The 2D cluster state on a square lattice can tolerate qubit loss up
to a rate $\sim 40\%$ \cite{Browne}. With a tolerable loss rate, a
2D graph-state network can be identified from the cluster state with
qubit loss, which can be converted to a new 2D cluster state on a
hexagonal lattice without qubit loss. The expected fraction of the
new cluster state $k(p_l)$ \cite{Browne}, i.e., the average length
of the path between nodes on the network is $1/k$, depends on the
loss rate $p_l$ (such a relation has been worked out numerically in
Ref.~\cite{Browne}). The errors on each path may affect the two
qubits, corresponding to the two connected nodes, on the final
hexagonal lattice. Therefore, on the new cluster state, the
probability of errors can be estimated as $p'_z \approx 3p_z/k$.
Here the factor $3$ is due to the three paths connected to each node
on the network. The thresholds for FTQC on the 2D cluster state are
more stringent than the thresholds for FTQC on one-dimensional
circuit architectures by a factor of approximately $10^2$
\cite{Raussendorf}. Because the one-dimensional-architecture
thresholds (for the circuit model) are approximately $10^{-5}$
\cite{Stephens08,Stephens09}, we thus use $10^{-7}$ as the
corresponding threshold of the 2D cluster-state model without qubit
loss to estimate the phase boundary for the transition in quantum
computational power. Therefore, the threshold of 2D models can be
estimated as
\begin{equation}
\label{eqn:2dplpz} p_z'\approx 10^{-7} \Rightarrow p_z \approx
\frac{1}{3}10^{-7}k(p_l).
\end{equation}
We numerically solve the temperature such that the above equation
holds to determine the `phase' boundary.
The resultant `phase
diagrams' for both models are shown in Fig.~\ref{fig:2Dphase}.

On 3D cluster states, one can encode quantum information with
topological codes, and hence error rates much higher than the 2D
threshold are tolerable. Without qubit loss, the error rate
threshold of 3D cluster states is $2.93\%$ for independent
phase-flip errors if the minimum-weight perfect matching algorithm
is used to find the likely distribution of errors.

{ On the 3D cluster state obtained from the construction in Fig.
\ref{fig:lattice2} (c), there are both independent errors and
correlated errors. By choosing the arrangement of particles as shown
in Fig. \ref{fig:lattice2} (d), the correlations occur between
errors either on directly connected qubits or two qubits oppositely
connected to the same qubit. The correlations of errors on directly
connected qubits can be neglected due to the error correction
algorithm, and the other type of correlations may affect the
threshold but not significantly \cite{RaussendorfAP,LiNJP}.}
Numerical evidence suggests that the threshold decreases
approximately linearly with the probability of qubit loss and it can
be tolerated up to $24.9\%$~\cite{Barrett}. As shown
in~\cite{Barrett}, the threshold of 3D models can therefore be
approximated as follows~\cite{Barrett}
\begin{equation}
\label{eqn:3dplpz} \frac{p_l}{24.9\%}+\frac{p_z}{2.93\%}\approx 1.
\end{equation}
Below this critical line, errors are correctable and the resource
state can be used for universal quantum computation. This relation
is then used to estimate the phase boundary for quantum
computational power, as shown in Fig.~\ref{fig:3Dphase}.

\section{Concluding remarks}
\label{sec:conclude} We have worked out the `phase diagrams' for the
quantum computational power for two different models in both two and
three dimensions. Our initial guess would be that such transition
might coincide with that in phases of matter~\cite{Darmawan}.
However, we find that instead quantum computational universality is
more intricate and may not persist at all points of a certain phase
of matter. The first model has a first-order phase transition at
$\delta=-2$ at zero temperature but no phase transitions at nonzero
temperatures. The isolated transition point does not locate at the
boundary in the quantum computational power. Said equivalently, in
this model the transition in the quantum computational power does
not coincide with the transition in phases of matter. Such a
non-coincidence was already hinted in Ref.~\cite{Darmawan}, where in
the quantum computational universality is likely to disappear at
certain point in the valence-bond solid phase. The second model does
not have a phase transition at all but has a transition in quantum
computational power at both zero and finite temperatures. The region
with quantum computational capability for both models survives to
higher temperatures in 3D than in 2D.

We also note that the ability to keep the interaction on while
performing quantum computation is not a general feature of the
model. It requires that all excited eigen-energies measured from the
ground energy must be rational relative to one
another~\cite{Thermal}. This only occurs when $\delta=d_z=0$, at
which the models possess the highest symmetry. Incidentally, the
closer to the symmetric point, the quantum computational power
appears to sustain to a higher temperature. The only other model
known to possess such a feature is actually the cluster-state model
itself~\cite{Oneway}.

{In our discussions of the FTQC thresholds, we have considered the
error sources come from the thermal effect, as we are interested in
the computational-power `phase diagram' of the states themselves. By
doing so we have assumed that measurements and other operations are
perfect.  These other errors can be included in the FTQC, and as
long as their error rates are small enough, they can also be
corrected by the error correction algorithm, but of course, will
reduce the tolerable temperature.}

\medskip \noindent {\bf Acknowledgment.} We thank J. Eisert for useful discussions. This work was supported by the National Science
Foundation under Grants No. PHY 1314748 and No. PHY 1333903 (TCW)
and by the National Research Foundation \& Ministry of Education of
Singapore (YL \& LCK).

\appendix
\begin{figure}
\vspace{1cm}
   \includegraphics[width=8cm]{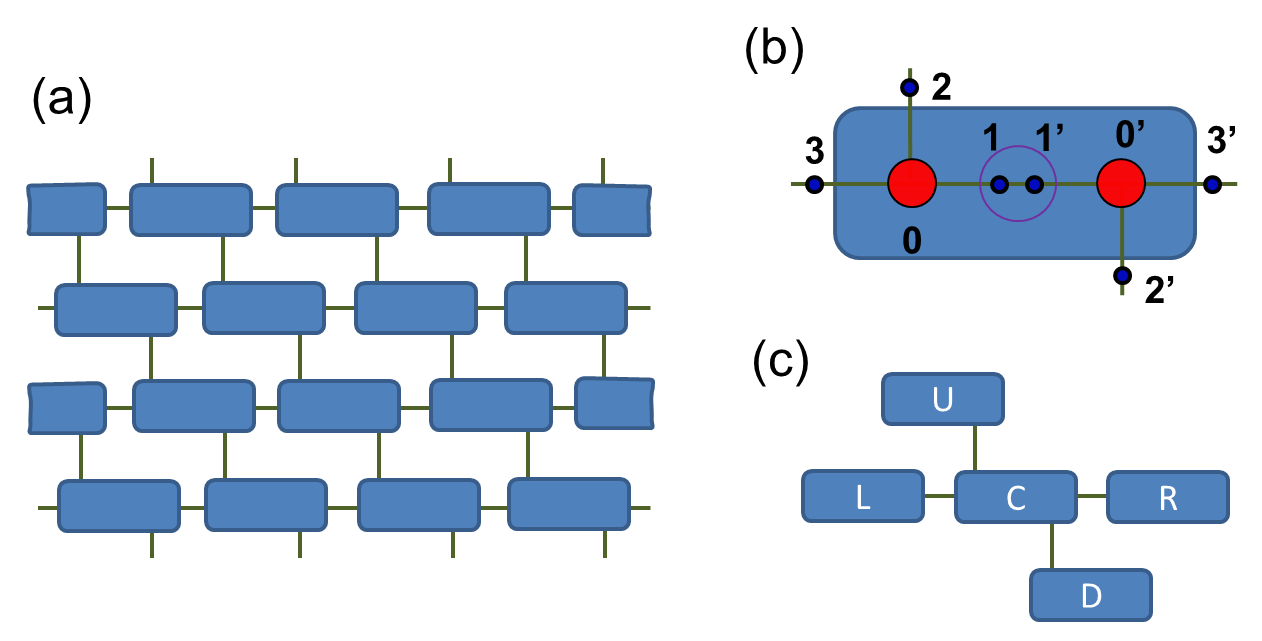}
  \caption{Generation of a cluster state on a square lattice: (a) original honeycomb (i.e. brickwall) lattice; (b) a block of spins that give rise to one logical cluster-state qubit; (c) a center logical block and its neighbor four blocks on a square lattice.
  \label{fig:honeysquare}}
\end{figure}
\section{Generation of cluster states and error analysis}
In this appendix we describe how the merging and CZ gates are
implemented by measuring bond particles. We also discuss the effect
of errors on qubits. To simplify notation, we will omit the overall
normalization. We assume that POVM's on all center particles have
been carried out and these particles become effectively qubits. We
illustrate how to obtain a cluster state on the square lattice, but
it is easily adapted to the bcc lattice.

(I) First let us consider how to merge two GHZ states of the form
$|0000\rangle+|1111\rangle$. This will be done by measuring the two
virtual qubits that form a bond particle. Denote other qubits not
involved by an underline, i.e.,
$|0000\rangle+|1111\rangle\rightarrow
|\underline{0}0\rangle+|\underline{1}1\rangle$. The two virtual
qubits will be measured in the basis
$\{|00\rangle\pm|11\rangle,|01\rangle\pm|10\rangle\}$, i.e., a
particular basis for the associated spin-3/2 bond particle. For
example, an outcome of $|00\rangle+|11\rangle$ will project the two
pairs of GHZ
$(|\underline{0}0\rangle+|\underline{1}1\rangle)(|\underline{0}0\rangle+|\underline{1}1\rangle)$
to $(|\underline{00}\rangle+|\underline{11}\rangle)$. Other outcomes
are equivalent to this up to a logical Pauli operation and translate
to basis change in the final cluster-state qubit.  The resulting
state is a six-qubit GHZ state: $|000000\rangle+|111111\rangle$.

(II) Second, to further shrink this to a five-qubit GHZ state we
measure one of the center spins in the basis of
$|\pm\rangle\equiv(|0\rangle\pm|1\rangle)/\sqrt{2}$, and the
resultant state for the remaining five qubits is
$|00000\rangle+|11111\rangle$, up to a logical Pauli Z correction,
which can be accounted for by a basis changed in later measurement
procedure.

\begin{table}[t]
\vspace{1cm}
  \begin{tabular}{|c|c|}
  \hline
   errors on 0-3 or 0'-3'
 & errors on  C, U, L, D, and R \\
 \hline
$X_0$ & $I$ (no error)\\
$X_{0'}$ & $X_C\equiv Z_UZ_LZ_DZ_R$\\
  $X_1$ or $X_{1'}$ & $Z_UZ_L$ \\
  $X_2$ & $Z_U$ \\
  $X_{2'}$ & $Z_D$ \\
 $X_3$ & $Z_L$\\
 $X_{3'}$ & $Z_R$\\
 $Z_i$ (for all $i$) & $Z_C$\\
 \hline
\end{tabular}
 \caption{\label{tbl:error} Consequence of qubit errors (from 0-3 and 0'-3' of block C in Fig.~\ref{fig:honeysquare})
 on logical cluster-state qubits. The logical error $X_C$ is equivalent to $Z_UZ_LZ_DZ_R$ due to the cluster-state stabilizer operator
 $X_CZ_UZ_LZ_DZ_R$. Note that the $Z_i$'s result in $Z_C$, and the probability of all such
 errors is already included in $2p_{Z_0}$.}
\end{table}

(III) Next we consider how to achieve the operation of CZ on two
center qubits of two GHZ pairs. This again is done by measuring an
associated bond particle (i.e. two virtual qubits) in a suitable
basis. The four basis states are ${\rm CZ} |\pm\pm\rangle$. It is
equivalent to applying a CZ gate between the two virtual qubits,
followed by a measurement in ${\pm\pm}$ basis. For illustration, we
again denote the two GHZ pairs by
$(|\underline{0}0\rangle+|\underline{1}1\rangle)(|\underline{0}0\rangle+|\underline{1}1\rangle)$.
The CZ operation between the two virtual qubits transforms the state
to
$(|\underline{0}0\rangle|\underline{0}0\rangle+|\underline{0}0\rangle|\underline{1}1\rangle+|\underline{1}1\rangle|\underline{0}0\rangle-|\underline{1}1\rangle|\underline{1}1\rangle)$.
Suppose $|++\rangle$ is obtained from measuring the two virtual
qubits (i.e. the bond particle), the remaining spins are projected
to
$(|\underline{0}\rangle|\underline{0}\rangle+|\underline{0}\rangle|\underline{1}\rangle+|\underline{1}\rangle|\underline{0}\rangle-|\underline{1}\rangle|\underline{1}\rangle)$,
i.e., a CZ gate has effectively applied between two center spins.

If all the bond particles are measured so as to induce CZ gates
between neighboring center spins, as in (III), then the center spins
will form a cluster-state on the original honeycomb lattice at the
end of the procedure. The consideration of faulty cluster state on
the honeycomb lattice could be carried out as done in the case of
the square lattice by Browne et al.~\cite{Browne} to extract the
corresponding ratio $k(p_l)$. But doing this is beyond the scope of
the present paper. Instead we use the result of $k(p_l)$ obtained in
Ref.~\cite{Browne} for the faulty square-lattice cluster state to
estimate the region where FTQC can be still be carried out. To do
this, we should aim to convert our spin network to a cluster state
on a square lattice. We note that although this may underestimate
the region of universality our goal is to show the existence of such
a region in both zero and non-zero temperatures.

To convert our original network of spins on the honeycomb lattice
(see e.g. Fig.~\ref{fig:honeysquare}) to form a cluster state on a
square lattice, we group two units of spin blocks as shown in
Fig.~\ref{fig:lattice2} to generate one logical qubit of a cluster
state. We label the spins as shown in Fig.~\ref{fig:honeysquare}c.
Virtual spins 1 and 1' are used to merge two GHZ states. Center spin
0 will be removed so as to shrink the 6-qubit GHZ to a 5-qubit GHZ
state. The remaining virtual qubits 2,3,2',3' will combine with
their partner virtual qubits to enact CZ gates on the center qubit
0' with neighboring such center qubits. The result will be a cluster
state on a square lattice. However, at finite temperatures thermal
errors occur and the result is a faulty cluster state. We thus
summarize the effect of (single-spin) errors on the logical qubits
of the cluster state in Table~\ref{tbl:error} for reference.

\end{document}